\documentclass[twocolumn,aps,pra]{revtex4-2}
\usepackage{bm,bbm}
\usepackage{amsfonts,amsmath,amssymb,latexsym}
\usepackage{graphicx}
\usepackage{mathtools}
\usepackage{xcolor}
\usepackage{hyperref}
\usepackage{physics}
\begin{document} 
\title{Polarization-controlled optical backflow in paraxial electromagnetic beams}
\author{Tomasz Rado\.zycki}
\email{t.radozycki@uksw.edu.pl}
\affiliation{Faculty of Mathematics and Natural Sciences, College of Sciences, Institute of Physical Sciences, Cardinal Stefan Wyszynski University in Warsaw, W\'oycickiego 1/3, 01-938 Warsaw, Poland} 

\begin{abstract}
Optical backflow in paraxial Gaussian beams is investigated within the Maxwell framework. Scalar potential representations are employed to identify conditions under which the longitudinal Poynting component becomes negative, showing that backflow is enabled by local suppression of the leading-order transverse field and the dominance of higher-order vectorial contributions. The spatial topology of backflow regions is shown to be governed by polarization through the number of independent local constraints on the transverse field. When the local polarization phase is free, as in the generic case of circular polarization, the leading-order field vanishes only at isolated points, giving rise to point-like backflow regions (extended curves may arise if an additional global phase constraint is imposed). In contrast, when the polarization phase is locally fixed, as for linear, radial, or azimuthal polarization, the suppression condition reduces to a single real constraint, resulting in extended backflow curves. Analytical Gaussian–polynomial solutions explicitly illustrate these effects. These results clarify the role of vectorial interference, establish a polarization-controlled backflow geometry, and provide a foundation for further studies of optical backflow in structured and nonparaxial beam configurations, as well as potential applications in optical manipulation and structured light design.
\end{abstract}

\maketitle

\section{Introduction}\label{intro}

Backflow is a counter-intuitive wave phenomenon in which the local energy or probability flux can oppose the direction suggested by the dominant momentum components of a wave. In quantum mechanics, the first systematic identification of such behaviour---where a wave packet composed only of positive momenta can nonetheless exhibit a negative probability current---was provided by Bracken and Melloy, who also quantified an upper bound on the magnitude of this effect \cite{brack}. Subsequent studies extended this analysis to relativistic and many-body systems, refining our understanding of the temporal and spectral properties of backflow \cite{mell,asfa,penz,mel2}. While historical investigations by Allcock in the late 1960s focused on the formal and interpretational aspects of the arrival-time problem in quantum mechanics, backflow appears there as a secondary implication of wave dynamics rather than the central subject of those works \cite{al1,al2,al3}.

In classical optics, an analog of quantum backflow emerges in structured electromagnetic fields, where the local energy flux need not follow the global direction of beam propagation. Suitably engineered optical fields can support spatially localized regions in which the longitudinal component of the Poynting vector becomes negative, even though the field propagates predominantly forward \cite{berry,saari}. This reversal of energy flow is a local interference effect arising from the vectorial structure of the electromagnetic field and cannot be interpreted as reflection or as the presence of an independently propagating backward wave. Consequently, optical backflow is absent in scalar theory and appears only when higher-order vectorial contributions are taken into account. Concrete realizations of this effect have been demonstrated in the focal region of tightly focused beams, including optical vortices and fields with polarization singularities, where nonparaxial corrections play a decisive role \cite{kot2018,kot2019}.

A key insight from earlier studies is that the magnitude and spatial structure of optical backflow depend sensitively on the beam geometry and polarization. In electromagnetic fields beyond the scalar regime, spatially localized regions with a negative longitudinal component of the Poynting vector have been demonstrated for various structured configurations, including superpositions of plane waves and space–time wave fields \cite{berry,saari}. In tightly focused beams both near-axis and off-axis reverse energy flux has been predicted, with the backflow appearing in regions where vectorial interference dominates \cite{kot2018,kot2019}. In the regime of paraxial vector Gaussian beams, exact analytical solutions of the paraxial Maxwell equations \cite{trvec} that retain higher-order terms in the small parameter $\varepsilon$ defined in (\ref{deeps}) show that selected polarization configurations exhibit negative longitudinal Poynting components, and that the magnitude of this backflow scales as $\varepsilon^4$ \cite{trback}. 

A commonly invoked conceptual criterion for backflow is that the wave packet be composed predominantly of positive spectral components. Gaussian beams do not strictly fulfill this condition, since their spectral decomposition contains small negative components. However, in the paraxial limit, the contribution of these negative spectral components to the longitudinal energy flux is exponentially suppressed, scaling as $\sim e^{-1/(2\varepsilon^2)}$. Therefore, this contribution is negligible compared to the power-law backflow effect $\sim\varepsilon^4$ captured by the full vectorial analysis. Because $\varepsilon$ is both small and tunable in typical experimental settings, the observable backflow associated with vectorial interference dominates over the tiny negative-spectrum contribution.

Backflow is not limited to simple Gaussian or plane-wave beams. Previous studies have demonstrated its occurrence in superpositions of waves, as well as in Bessel and exponential beams, and even in relativistic and gravitational wave contexts \cite{ibba,ibbc}. Interference-induced regions of reversed Poynting flux have been demonstrated in a variety of settings, including direct measurements of reverse energy flow in the tight focal spot of vortex beams \cite{kot2020exp}, complex azimuthal backflow patterns in multi-beam interference \cite{he2026azim}, and single-photon azimuthal backflow probed via weak measurement \cite{zhang2025sp}. These findings confirm the practical relevance of backflow beyond purely theoretical models and point toward potential applications in precision optical manipulation of particles and nanoparticles using optical tweezers \cite{ash,chu,miller,pad,kolbow,gaon}. Additionally, backflow can be exploited for tailoring complex energy flows in structured light and designing advanced photonic devices, including sub-wavelength imaging systems and optical metrology setups \cite{luo,neice,cheben,ren,kaz}. By exploiting localized backflow regions, it becomes possible to engineer energy transport pathways and optical forces at the microscale and nanoscale, enabling new functionalities in optical trapping, beam shaping, and high‑resolution optical instrumentation.

In this work, optical backflow in fundamental Gaussian beams has been elucidated within the exact paraxial Maxwell framework, with polarization-dependent effects explicitly taken into account. Analytical solutions, expressed in terms of scalar potentials, have been employed to systematically investigate the conditions under which a localized reversal of the longitudinal Poynting vector occurs (Sec.~\ref{pmf} and Sec.~\ref{szanalysis}). Following \cite{trvec}, the solutions of the full paraxial Maxwell equations are introduced in Sec.~\ref{pmf}, while the spectral contributions to momentum and their relative suppression are analyzed in Sec.~\ref{szanalysis}. The polarization-dependent geometry of optical backflow is further explored (Sec.~\ref{lbm}), and concrete examples, supported by graphical illustrations, are presented in Sec.~\ref{ce}, demonstrating configurations that are, in principle, experimentally accessible.

\section{Paraxial Electromagnetic Fields}\label{pmf}

For the present analysis, only the \emph{sign} of the longitudinal ($z$-) component of the Poynting vector is relevant, rather than its absolute magnitude or the particular system of units employed. This allows us to simplify the formalism by introducing dimensionless variables. Specifically, we define
\begin{equation}\label{resca}
\boldsymbol{\xi} = k \boldsymbol{r}, \qquad \tau = \omega t,
\end{equation}
where $k=\omega/c$ is the wavenumber, and work with suitably normalized, dimensionless electric and magnetic fields. This rescaling leaves the direction---and hence the sign---of the Poynting vector unchanged, while significantly simplifying the expressions used in the subsequent analysis.

In what follows, all physical quantities are understood as being averaged over the rapid temporal oscillations of the electromagnetic field, i.e., over the dimensionless time variable $\tau$. As a consequence, $\tau$ does not appear explicitly in the expressions below. Furthermore, dimensionless electromagnetic fields can be related to physical (dimensional) fields by factoring out constants $E_0$ and $B_0$ (which can be chosen real and are inessential to the present analysis) associated with the overall beam intensity:
\begin{equation}\label{polebe}
\bm{E} = E_0 \bm{\mathcal{E}}, \qquad \bm{B} = B_0 \bm{\mathcal{B}}.
\end{equation}

The time-averaged longitudinal component of the (dimensionless) Poynting vector then takes the form
\begin{equation}\label{poyz}
\mathcal{S}_z=\frac{1}{2}\,\Re\!\left(\mathcal{E}_x \mathcal{B}_y^{*}-\mathcal{E}_y \mathcal{B}_x^{*}\right),
\end{equation}
where the asterisk denotes complex conjugation. Only the transverse field components contribute to the energy flow along the propagation direction.

Within the maximal paraxial approximation, the electric field components are given by~\cite{trvec}
\begin{subequations}\label{parpeeq}
\begin{align}
&\mathcal{E}_{x}=e^{i\xi_z}\Big[\partial_{\xi_x}\Big(1-\delta^2\,\frac{i}{2}\,\partial_{\xi_z} \Big)V_+\nonumber\\
&\qquad-i\partial_{\xi_y}\Big(1+\delta^2\,\frac{i}{2}\,\partial_{\xi_z}\Big)V_-\Big],\label{parpexe}\\
&\mathcal{E}_{y}=e^{i\xi_z}\Big[i\partial_{\xi_x}\Big(1+\delta^2\,\frac{i}{2}\,\partial_{\xi_z} \Big)V_-\nonumber\\
&\qquad+\partial_{\xi_y}\Big(1-\delta^2\,\frac{i}{2}\,\partial_{\xi_z}\Big)V_+\Big],\label{parpeye}\\
&\mathcal{E}_{z}=2\delta\,e^{i\xi_z}\partial_{\xi_z}\widetilde{V}_+,\label{parpeze}
\end{align}
\end{subequations}
and the magnetic field components read
\begin{subequations}\label{parpbeq}
\begin{align}
&\mathcal{B}_{x}=e^{i\xi_z}\Big[-i\partial_{\xi_x}\Big(1-\delta^2\,\frac{i}{2}\,\partial_{\xi_z} \Big)V_-\nonumber\\
&\qquad-\partial_{\xi_y}\Big(1+\delta^2\,\frac{i}{2}\,\partial_{\xi_z}\Big)V_+\Big],\label{parpbxe}\\
&\mathcal{B}_{y}=e^{i\xi_z}\Big[\partial_{\xi_x}\Big(1+\delta^2\,\frac{i}{2}\,\partial_{\xi_z} \Big)V_+\nonumber\\
&\qquad-i\partial_{\xi_y}\Big(1-\delta^2\,\frac{i}{2}\,\partial_{\xi_z}\Big)V_-\Big],\label{parpbye}\\
&\mathcal{B}_{z}=-2i\delta\, e^{i\xi_z}\partial_{\xi_z}\widetilde{V}_-.\label{parpbze}
\end{align}
\end{subequations}
Here and in the following we use the shorthand notation $\partial_{\xi_j} \equiv \partial / \partial \xi_j$ for $j = x,y,z$. 

The parameter $\delta$ formally equals unity and carries no independent physical meaning; it is retained as an auxiliary marker that distinguishes the leading paraxial contributions from higher-order correction terms required for the exact
satisfaction of the paraxial Maxwell equations.  It will play a key role in identifying subsequent contributions to the Poynting vector that involve small derivatives with respect to $\xi_z$. The different powers of $\delta$ reflect the paraxial hierarchy of the field components: the longitudinal field appears at first order, whereas corrections
to the transverse components enter at the next order.

The scalar potentials $V_\pm$ satisfy the paraxial wave equation
\begin{equation}\label{scpaw}
\left(\triangle_\perp+2 i\,\partial_{\xi_z}\right)V_\pm = 0,
\end{equation}
where
\begin{equation}
\triangle_\perp=\partial_{\xi_x}^{2}+\partial_{\xi_y}^{2}
\end{equation}
denotes the transverse Laplacian. 

As is well known, beams with a Gaussian-like transverse profile---such as Laguerre-Gaussian (LG) and Bessel-Gaussian (BG) modes---are the most typical examples of axisymmetric paraxial beams. They are characterized by two fundamental length scales: $w_0$, the radius of the beam waist in the transverse plane, and the Rayleigh range along the propagation axis, $z_R = \frac{1}{2} k w_0^2$. In typical situations, they satisfy the inequalities
\begin{equation}\label{defyb}
\lambda \ll w_0 \ll z_R,
\end{equation}
which naturally motivates the introduction of the small, dimensionless paraxial parameter
\begin{equation}\label{deeps}
\varepsilon = \frac{1}{k w_0} = \frac{\lambda}{2\pi w_0} = \frac{w_0}{2 z_R}.
\end{equation}
In most practical situations, the parameter $\varepsilon$ is very small, around $10^{-4}$. However, for tightly focused beams \cite{neice,luo,cheben,ren,kaz,gaon,kolbow,sharma,sted}, it can increase significantly, reaching values of approximately $0.2$–$0.3$. While some paraxial beam families, such as gamma beams \cite{trgam}, do not follow Gaussian-like profiles, the general framework presented here remains applicable to these cases as well.

Terms containing derivatives with respect to the longitudinal coordinate $\xi_z$ in the expressions for the electric and magnetic fields (\ref{parpeeq}) and (\ref{parpbeq}) are of order $\varepsilon^2$. These terms are frequently neglected in the standard paraxial approximation~\cite{patta,es,hil95,que,nomoto,pastor}; however, as will be discussed in the next section, they play a fundamental role in determining the local sign of the longitudinal Poynting vector and in the emergence of effects such as optical backflow. This conclusion will be further confirmed by concrete examples in Section~\ref{ce}.

\section{Decomposition of the longitudinal Poynting vector}\label{szanalysis}

To analyze optical backflow, the longitudinal component of the Poynting vector, $\mathcal S_z$, can be naturally decomposed according to the structure of the paraxial expansion into
\begin{enumerate}
\item $\mathcal S_z^{(0)}$: terms involving only transverse derivatives $(\partial_{\xi_x},\partial_{\xi_y})$, corresponding to order $\delta^{0}$, 
\item $\mathcal S_z^{(1)}$: mixed terms containing one transverse and one
longitudinal derivative $(\partial_{\xi_z})$, corresponding to order $\delta^{2}$, 
\item $\mathcal S_z^{(2)}$: terms involving only longitudinal derivatives $(\partial_{\xi_z})$, corresponding to order $\delta^{4}$.
\end{enumerate}

It should be noted that this decomposition is unrelated to the spin–orbital separation of electromagnetic momentum commonly discussed in the literature \cite{bek}, as it arises purely from the hierarchical structure of the paraxial expansion rather than from a redistribution of the total momentum density.

\subsection{Explicit forms}

Let us now use the paraxial field expressions (\ref{parpeeq})--(\ref{parpbeq}) to decompose the longitudinal Poynting vector into the contributions as said above. 

The \emph{zeroth-order} term, containing only transverse derivatives, reads
\begin{align}\label{sz0fulla}
\mathcal{S}_z^{(0)} &= \frac{1}{2} \Re \Big[
(\partial_{\xi_x}V_+ - i \partial_{\xi_y} V_-)(\partial_{\xi_x} V_+^* + i \partial_{\xi_y} V_-^*) \nonumber\\
&\quad - (i \partial_{\xi_x} V_- + \partial_{\xi_y} V_+)(i \partial_{\xi_x} V_-^* - \partial_{\xi_y} V_+^*) 
\Big]
\end{align}
and can be rewritten in the useful form
\begin{equation}\label{sz0full}
\mathcal{S}_z^{(0)}= \frac{1}{2} \Big[|\partial_{\xi_x}V_+ - i \partial_{\xi_y} V_-|^2 + |\partial_{\xi_y} V_+ + i \partial_{\xi_x} V_-|^2\Big] \ge 0.
\end{equation}

Obviously, this term is always nonnegative and represents the leading-order longitudinal energy flow. Its vanishing is a necessary prerequisite for backflow to occur, as any nonzero $\mathcal{S}_z^{(0)}$ would dominate over higher-order contributions.

The \emph{first-order} contribution, proportional to $\delta^{2}$, contains one longitudinal derivative and reads
\begin{align}\label{sz1full}
\mathcal{S}_z^{(1)} &= \frac{1}{2} \Re \Big[\frac{i}{2} (\partial_{\xi_x}V_+ - i \partial_{\xi_y} V_-)\,
\partial_{\xi_z} (-\partial_{\xi_x}V_+^* + i \partial_{\xi_y}V_-^*) \nonumber\\
&\quad - \frac{i}{2}\, \partial_{\xi_z} (\partial_{\xi_x}V_+ + i \partial_{\xi_y} V_-)\, (\partial_{\xi_x} V_+^* + i \partial_{\xi_y} V_-^*) \nonumber\\
&\quad - \frac{i}{2} (i\partial_{\xi_x} V_- + \partial_{\xi_y} V_+)\,\partial_{\xi_z} (i \partial_{\xi_x} V_-^* + \partial_{\xi_y} V_+^*) \nonumber\\
&\quad - \frac{i}{2}\, \partial_{\xi_z} (i \partial_{\xi_x} V_- - \partial_{\xi_y} V_+)\, (i \partial_{\xi_x} V_-^* - \partial_{\xi_y} V_+^*)\Big].
\end{align}
After isolating this contribution, the auxiliary parameter $\delta$ has been set to unity, as discussed earlier.
A straightforward calculation shows that the expression inside the brackets is purely imaginary. Therefore, its real part vanishes identically:
\begin{align}\label{sz1simpl}
\mathcal{S}_z^{(1)} &= - \frac{1}{4} \partial_{\xi_z} \Re \Big[ i \big(|\partial_{\xi_x}V_+|^2 + |\partial_{\xi_y} V_+|^2 \nonumber\\
&\quad - |\partial_{\xi_x} V_-|^2 - |\partial_{\xi_y} V_-|^2 \big) \Big] = 0.
\end{align}

Finally, the \emph{second-order} contribution, proportional to $\delta^{4}$ and involving only longitudinal derivatives, takes the form
\begin{align}\label{sz2fulla}
\mathcal{S}_z^{(2)} &= - \frac{1}{8} \Re \Big[\partial_{\xi_z} (\partial_{\xi_x}V_+ + i \partial_{\xi_y} V_-)\,
\partial_{\xi_z} (\partial_{\xi_x} V_+^* - i \partial_{\xi_y} V_-^*) \nonumber\\
&\quad - \partial_{\xi_z} (i \partial_{\xi_x} V_- - \partial_{\xi_y} V_+)\,\partial_{\xi_z} (i \partial_{\xi_x} V_-^* + \partial_{\xi_y} V_+^*)\Big] \nonumber\\
&= -\frac{1}{8} \Big(|\bm{\nabla}_\perp \partial_{\xi_z} V_+|^2 + |\bm{\nabla}_\perp \partial_{\xi_z} V_-|^2
\Big) \nonumber\\
&\quad - \frac{1}{4} \Im\!\left(\bm{\nabla}_\perp \partial_{\xi_z} V_+\times\bm{\nabla}_\perp \partial_{\xi_z} V_-^*
\right), 
\end{align}
and can be arranged to make it clear that the expression is non-positive:
\begin{align}\label{sz2full}
\mathcal{S}_z^{(2)} &= -\frac{1}{8} \Big[|\partial_{\xi_x}\partial_{\xi_z} V_+ + i \partial_{\xi_y}\partial_{\xi_z} V_-|^2 \nonumber\\
&\quad + |\partial_{\xi_y}\partial_{\xi_z} V_+ - i \partial_{\xi_x}\partial_{\xi_z} V_-|^2\Big] \le 0 .
\end{align}
This term is solely responsible for any negative contributions to $\mathcal{S}_z$, i.e., for optical backflow.

Summing all contributions, the total longitudinal Poynting vector reads
\begin{align}\label{szfullfinal}
\mathcal{S}_z &= \mathcal{S}_z^{(0)} + \mathcal{S}_z^{(1)} + \mathcal{S}_z^{(2)} \nonumber\\
&= \frac{1}{2} \Big[|\partial_{\xi_x}V_+ - i \partial_{\xi_y} V_-|^2 + |\partial_{\xi_y} V_+ + i \partial_{\xi_x} V_-|^2\Big] \nonumber\\
&\quad - \frac{1}{8} \Big[|\partial_{\xi_x}\partial_{\xi_z} V_+ + i \partial_{\xi_y}\partial_{\xi_z} V_-|^2 \nonumber\\
&\quad + |\partial_{\xi_y}\partial_{\xi_z} V_+ - i \partial_{\xi_x}\partial_{\xi_z} V_-|^2 \Big].
\end{align}

The analysis above demonstrates that optical backflow can occur only in regions where the zeroth-order, nonnegative contributions vanish or are strongly suppressed, allowing the negative fourth-order terms to dominate locally.  

\subsection{Implications for backflow}

The vanishing of the leading-order term $\mathcal{S}_z^{(0)}$ requires satisfying the conditions
\begin{equation}\label{szzero}
\partial_{\xi_x} V_+ = i \, \partial_{\xi_y} V_-,\qquad
\partial_{\xi_y} V_+ = - i \, \partial_{\xi_x} V_-.
\end{equation}

This is not merely a convenient simplification. If $\mathcal{S}_z^{(0)} \neq 0$, the corresponding electric and magnetic fields remain finite, and the longitudinal energy flow is strictly positive. Hence, the vanishing of $\mathcal{S}_z^{(0)}$ is a natural and necessary condition for the occurrence of optical backflow. 

However, it can be readily verified that enforcing conditions (\ref{szzero}) throughout extended regions would be excessively restrictive. In such domains, the electromagnetic fields reduce to the residual, higher-order contributions:
\begin{subequations}\label{szb}
\begin{align}
\mathcal{E}_x &\approx - i\delta^2 \, e^{i\xi_z} \, \partial_{\xi_x} \partial_{\xi_z} V_+,\\
\mathcal{E}_y &\approx - i\delta^2 \, e^{i\xi_z} \, \partial_{\xi_y} \partial_{\xi_z} V_+,\\
\mathcal{E}_z &\approx 2\delta \, e^{i\xi_z} \, \partial_{\xi_z} V_+,\\ 
\mathcal{B}_x &\approx - i\delta^2 \, e^{i\xi_z} \, \partial_{\xi_y} \partial_{\xi_z} V_+,\\
\mathcal{B}_y &\approx i\delta^2 \, e^{i\xi_z} \, \partial_{\xi_x} \partial_{\xi_z} V_+,\\
\mathcal{B}_z &\approx - 2 i\delta \, e^{i\xi_z} \, \partial_{\xi_z} V_-,
\end{align}
\end{subequations}
which result from the cancellation of the leading-order transverse terms.

Mathematically, the conditions (\ref{szzero}) resemble Cauchy-Riemann relations and as such directly imply $\triangle_\perp V_\pm = 0$. Since the scalar potentials also satisfy the paraxial equation (\ref{scpaw}), this would enforce $\partial_{\xi_z} V_\pm = 0$, leading to identically vanishing fields, $\bm{\mathcal{E}} = \bm{\mathcal{B}} = 0$, and a zero value of the longitudinal Poynting vector. This demonstrates that imposing (\ref{szzero}) across extended regions is excessively stringent, which would preclude the occurrence of backflow.

Instead, it is sufficient for these conditions to hold only at isolated points. By continuity of the fields, small neighborhoods exist around such points where the zeroth-order contribution $\mathcal{S}_z^{(0)}$ is suppressed. In these neighborhoods, the negative fourth-order term $\mathcal{S}_z^{(2)}$ can locally dominate, giving rise to optical backflow. Concrete examples and further illustrations of this mechanism will be presented in the following section.

Summarizing, the optical backflow arises when
\begin{subequations}\label{condpm}
\begin{align}
&\partial_{\xi_x} V_+ - i \, \partial_{\xi_y} V_-=0, \quad \mathrm{and}\label{condpm1}\\
&\partial_{\xi_y} V_+ + i \, \partial_{\xi_x} V_-=0,\label{condpm2}
\end{align}
\end{subequations}
 at given points, while 
\begin{subequations}\label{condpma}
\begin{align}
&\partial_{\xi_x}\partial_{\xi_z} V_+ + i \, \partial_{\xi_y}\partial_{\xi_z} V_-\neq 0, \quad\mathrm{or}\label{condpma1}\\
&\partial_{\xi_y}\partial_{\xi_z} V_+ -  i \, \partial_{\xi_x}\partial_{\xi_z} V_-\neq 0\label{condpma2}
\end{align}
\end{subequations}
at the same points.

In fact, conditions (\ref{condpm}) state that the main-order transverse components of the electric field vanish at a given point, $\mathcal{E}_{x}^{(0)}=\mathcal{E}_{y}^{(0)}=0$. This can be inferred directly from (\ref{parpeeq}) or derived from the paraxial Maxwell equations \cite{trvec}. The local suppression of the dominant field components allows paraxial corrections to become significant. Furthermore, conditions (\ref{condpma}) ensure that not only do the dominant transverse components vanish, but local field variations along the propagation axis exhibit nontrivial higher-order derivatives. Such configurations typically occur near phase singularities or in regions of rapid field-structure variation, as found in spatially superoscillatory and interference-dominated wave fields \cite{dj1,cheche}. In particular, spatial superoscillatory fields exhibit isolated or extended regions where the dominant transverse field components are locally suppressed, directly satisfying the conditions for optical backflow described above, and these regions, where higher-order gradients prevail, can be structurally robust even in the absence of precise adjustment of beam parameters.

\section{Potentials, helicity and polarization}\label{hel}

The scalar potentials $V_\pm$ do not correspond directly to helicity states.  
In particular, setting $V_-=0$ or $V_+=0$ does \emph{not} produce states with $\mathcal{F}_+=0$ or $\mathcal{F}_-=0$ (where $\bm{\mathcal F}_\pm = \bm{\mathcal E} \pm i \bm{\mathcal B}$ are the Riemann-Silberstein vectors), not even approximately.  
The helicity operator $\bm{\nabla}_\xi \times$ acts as
\begin{equation}\label{helop}
\bm{\nabla}_\xi \times \bm{\mathcal{F}}_\pm = \pm \bm{\mathcal{F}}_\pm + \mathcal{O}(\partial_{\xi_z}^2\bm{\mathcal{F}}),
\end{equation}
regardless of whether $V_+$ or $V_-$ is set to zero.  
Here and below, $\mathcal{O}(\partial_{\xi_z}^2 \bm{\mathcal{F}})$ denotes terms of order $\varepsilon^4$ in the paraxial expansion, since each longitudinal derivative $\partial_{\xi_z}$ scales as $\varepsilon^2$.

It is convenient to define the combinations
\begin{equation}\label{vrl}
V_R = \frac{1}{2}\left(V_+ + V_-\right), \qquad V_L = \frac{1}{2}\left(V_+ - V_-\right),
\end{equation}
which allow for a direct identification of the approximate helicity components. Explicitly, setting $V_L = 0$ gives
\begin{equation}\label{flzero}
\bm{\mathcal{F}}_- = 0, 
\end{equation}
while the positive-helicity RS vector satisfies
\begin{equation}\label{curlfp}
\bm{\nabla}_\xi \times \bm{\mathcal{F}}_+ = \bm{\mathcal{F}}_+ + \mathcal{O}(\partial_{\xi_z}^2 \bm{\mathcal{F}}).
\end{equation}
Conversely, $V_R = 0$ yields
\begin{equation}\label{frzero}
\bm{\mathcal{F}}_+ = 0, \qquad \bm{\nabla}_\xi \times \bm{\mathcal{F}}_- = -\bm{\mathcal{F}}_- + \mathcal{O}(\partial_{\xi_z}^2 \bm{\mathcal{F}}).
\end{equation}

Using the complex variables
\begin{equation}\label{compva}
\zeta = \xi_x + i \xi_y, \qquad \bar\zeta = \xi_x - i \xi_y,
\end{equation}
the RS vectors can be expressed in a compact way. For positive helicity:
\begin{subequations}\label{rsp}
\begin{align}
\mathcal{F}_{+x} &= 4\, e^{i\xi_z} \Big[\partial_\zeta -\delta^2 \frac{i}{2}\,\partial_{\bar{\zeta}}\partial_{\xi_z}\Big] V_R,\\
\mathcal{F}_{+y} &= 4\, e^{i\xi_z} \Big[i \partial_\zeta -\delta^2 \frac{1}{2}\,\partial_{\bar{\zeta}}\partial_{\xi_z}\Big] V_R,\;\;\;\;\;\\ \mathcal{F}_{+z} &= 4\delta\, e^{i\xi_z} \partial_{\xi_z} V_R,
\end{align}
\end{subequations}
and for negative helicity:
\begin{subequations}\label{rspm}
\begin{align}
\mathcal{F}_{-x} &= 4\, e^{i\xi_z} \Big[\partial_{\bar\zeta} -\delta^2 \frac{i}{2}\,\partial_{\zeta}\partial_{\xi_z}\Big] V_L,\\
\mathcal{F}_{-y} &= 4\, e^{i\xi_z} \Big[-i \partial_{\bar\zeta} +\delta^2 \frac{1}{2}\,\partial_{\zeta}\partial_{\xi_z}\Big] V_L,\\
\mathcal{F}_{-z} &= 4\delta\, e^{i\xi_z} \partial_{\xi_z} V_L.
\end{align}
\end{subequations}
These expressions explicitly demonstrate Eqs.~(\ref{flzero}) and (\ref{frzero}). The R/L labeling convention merely serves as a notation for the helicity components and does not affect the analysis of optical backflow, which depends solely on the sign of the longitudinal ($z$-) component of the Poynting vector.

Helicity emerges from the field constructed from $V_R$ and $V_L$, rather than from the individual potentials $V_\pm$.  
Polarization describes the temporal evolution of the electric field at a fixed spatial point; for monochromatic fields, this is fully captured by the relative amplitudes and phases of the complex components.

\subsection*{Circular polarization}

To leading paraxial order, the scalar potentials $V_R$ and $V_L$ correspond to the right- and left-handed circular polarization components, respectively. Without loss of generality, let us set 
\begin{equation}\label{vlze} 
V_L = 0 , \qquad V_R \neq 0. 
\end{equation} 
Equation~(\ref{rspm}) then implies $\bm{\mathcal{E}}=i\bm{\mathcal{B}}$, and using (\ref{rsp}) one obtains 
\begin{subequations}\label{rspe} 
\begin{align} 
&\mathcal{E}_{x}=2\,e^{i\xi_z}\left[\partial_{\zeta}-\delta^2\frac{i}{2}\,\partial_{\bar{\zeta}}\partial_{\xi_z}\right]V_R,\label{rspex}\\ 
&\mathcal{E}_{y}=2\,e^{i\xi_z}\left[i\partial_{\zeta}-\delta^2\frac{1}{2}\,\partial_{\bar{\zeta}}\partial_{\xi_z}\right]V_R,\label{rspey}\\ 
&\mathcal{E}_{z}=2\delta\,e^{i\xi_z}\partial_{\xi_z}V_R.\label{rspez} 
\end{align} 
\end{subequations} 
Consequently (up to the leading order), 
\begin{equation}\label{proe} 
\mathcal{E}_{y}=i\mathcal{E}_{x}. 
\end{equation} 
This relation corresponds to right-handed circular polarization (in the optical convention). Indeed, the transverse components of the real electric field at a given point satisfy
\begin{subequations}\label{trele} 
\begin{align} 
E_x&=\Re\!\left(e^{-i\tau}\mathcal{E}_{x}\right) =\mathcal{E}_{rx}\cos\tau+\mathcal{E}_{ix}\sin\tau\nonumber\\ 
&=\sqrt{\mathcal{E}_{rx}^2+\mathcal{E}_{ix}^2}\cos(\tau-\alpha),\label{trelex}\\ 
E_y&=\Re\!\left(ie^{-i\tau}\mathcal{E}_{x}\right) =\mathcal{E}_{rx}\sin\tau-\mathcal{E}_{ix}\cos\tau\nonumber\\ &=\sqrt{\mathcal{E}_{rx}^2+\mathcal{E}_{ix}^2}\sin(\tau-\alpha),\label{treley} 
\end{align} 
\end{subequations} 
where the subscripts $r$ and $i$ denote the real and imaginary parts of the corresponding complex field components and 
\begin{equation}\label{defisc} 
\cos\alpha=\frac{\mathcal{E}_{rx}}{\sqrt{\mathcal{E}_{rx}^2+\mathcal{E}_{ix}^2}},\qquad \sin\alpha=\frac{\mathcal{E}_{ix}}{\sqrt{\mathcal{E}_{rx}^2+\mathcal{E}_{ix}^2}}. 
\end{equation} 
These expressions describe a vector rotating clockwise in the transverse plane when viewed from the wavefront looking back toward the source. 

The opposite sense of rotation is obtained when $V_R=0$ and $V_L\neq 0$. In this case one gets (again up to the leading order)
\begin{equation}\label{proea} 
\mathcal{E}_{y}=-i\mathcal{E}_{x}, 
\end{equation} 
and 
\begin{subequations}\label{trelea} 
\begin{align} 
E_x&=\Re\!\left(e^{-i\tau}\mathcal{E}_{x}\right) =\mathcal{E}_{rx}\cos\tau+\mathcal{E}_{ix}\sin\tau\nonumber\\ 
&=\sqrt{\mathcal{E}_{rx}^2+\mathcal{E}_{ix}^2}\cos(\tau-\alpha),\label{treleax}\\ 
E_y&=\Re\!\left(-ie^{-i\tau}\mathcal{E}_{x}\right) =-\mathcal{E}_{rx}\sin\tau+\mathcal{E}_{ix}\cos\tau\nonumber\\ 
&=-\sqrt{\mathcal{E}_{rx}^2+\mathcal{E}_{ix}^2}\sin(\tau-\alpha),\label{treleay} 
\end{align} 
\end{subequations} 

Although the expressions~(\ref{trele}) and~(\ref{trelea}) are formally analogous, the quantities denoted by $\mathcal{E}_{rx}$, $\mathcal{E}_{ix}$, etc., do not represent identical field amplitudes in the two cases. This is because the electric (and magnetic) fields are generated from the scalar potentials $V_R$ and $V_L$ through distinct differential relations, even though they have the same functional form. Paraxial corrections of order $\varepsilon^2$ lead to small local deviations from exact circular polarization, which may be interpreted as a weak ellipticity of the field.

\subsection*{Linear polarization}

Linear polarization arises when the electric field oscillates along a fixed direction in the transverse plane. In contrast to the circularly polarized case, linear polarization is not associated with setting one of the potentials to zero. Instead, it requires a specific correlation between the scalar potentials $V_+$ and $V_-$. From the paraxial expressions for the electric field components (\ref{parpeeq}), it follows that, to leading order in $\varepsilon$, the electric field is linearly polarized along the $x$ axis if the scalar potentials obey 
\begin{equation}\label{conlix} 
\partial_{\xi_x} V_- = i\,\partial_{\xi_y} V_+ . 
\end{equation} 
Under this constraint, the leading transverse contribution to $\mathcal{E}_y$ cancels, and the electric field is aligned predominantly along the $x$ direction, with only small paraxial corrections of order $\varepsilon^2$.  Similarly, linear polarization along the $y$ axis is obtained when the leading contribution to $\mathcal{E}_x$ vanishes, which requires 
\begin{equation}\label{conliy} 
\partial_{\xi_y} V_- = - i\,\partial_{\xi_x} V_+ . 
\end{equation} 
In this latter case, the transverse electric field oscillates predominantly along the $y$ direction, again up to corrections of order $\varepsilon^2$. 

One can summarize that, within the paraxial framework, linear polarization emerges from specific derivative relations between the scalar potentials $V_+$ and $V_-$, rather than from a simple suppression of one of them. As in the circularly polarized case, higher-order paraxial terms lead to small local deviations from ideal linear polarization.

\subsection*{Radial polarization}

For axisymmetric beams, radial polarization is defined by the vanishing of the azimuthal field component,
\begin{equation}\label{znif}
\mathcal{E}_\phi = 0,
\end{equation}
across a finite region. It is convenient, alongside Cartesian coordinates $\xi_x$ and $\xi_y$, to introduce polar coordinates:
\begin{equation}
\xi = \sqrt{\xi_x^2 + \xi_y^2}, \qquad \phi = \arctan\frac{\xi_y}{\xi_x}.
\end{equation}

Using leading-order paraxial expressions,
\begin{subequations}
\begin{align}
\mathcal{E}_x &\simeq e^{i\xi_z}\left(\partial_{\xi_x} V_+ - i\,\partial_{\xi_y} V_-\right),\\
\mathcal{E}_y &\simeq e^{i\xi_z}\left(i\,\partial_{\xi_x} V_- + \partial_{\xi_y} V_+\right),
\end{align}
\end{subequations}
the azimuthal component reads
\begin{equation}
\mathcal{E}_\phi = -\mathcal{E}_x \,\sin\phi  +\mathcal{E}_y \, \cos\phi ,
\end{equation}
and condition (\ref{znif}) in terms of the potentials becomes
\begin{equation}\label{radpol}
i \, \partial_\xi V_- + \frac{1}{\xi} \, \partial_\phi V_+ = 0.
\end{equation}

\subsection*{Azimuthal polarization}

Azimuthal polarization is defined by the vanishing of the radial component of the transverse electric field,
\begin{equation}
\mathcal{E}_\xi = 0 ,
\end{equation}
which implies that the field oscillates tangentially to circles centered on the propagation axis.
In Cartesian components this condition reads
\begin{equation}
\mathcal{E}_x \,\cos\phi  + \mathcal{E}_y \,\sin\phi  = 0 .
\end{equation}

Using the leading-order paraxial expressions for $\mathcal{E}_x$ and $\mathcal{E}_y$ and transforming to polar coordinates in the transverse plane, this requirement can be written in terms of the scalar potentials as
\begin{equation}\label{azpol}
i\partial_\xi V_+ + \frac{1}{\xi}\,\partial_\phi V_- = 0 .
\end{equation}
Thus, azimuthal polarization results from a specific coupling between the radial and azimuthal derivatives of the scalar potentials, similarly to the case of radial polarization.
More generally, in wave fields with complex spatial structure---such as modal superpositions, topological beams, or fields with phase or polarization singularities---the local annihilation of transverse components has a clear physical basis. Superoscillatory models demonstrate that certain regions of the field may exhibit rapid phase gradients and amplitude zeros even when the global spectral content is bounded \cite{dj1}. Such regions naturally satisfy conditions (\ref{condpm}) and (\ref{condpma}), facilitating optical backflow without the need for artificial parameter adjustment.

\section{Polarization-dependent geometry of optical backflow}\label{lbm}

The emergence of optical backflow is governed by the local structure of the transverse electric field.
As shown in the previous section, the longitudinal component of the Poynting vector can become negative only at points ($\bm{\xi}_0$) where the leading-order contribution vanishes,
\begin{equation}\label{lbm0}
\mathcal S_z^{(0)}(\bm{\xi}_0)=0,
\end{equation}
which is equivalent to the local condition
\begin{equation}\label{lbm1}
\mathcal E_x^{(0)}(\bm{\xi}_0)=
\mathcal E_y^{(0)}(\bm{\xi}_0)=0 .
\end{equation}
The spatial structure of the set of points $\bm{\xi}_0$ satisfying \eqref{lbm1} depends crucially on the polarization state of the field. The following considerations are of a general nature and do not preclude special choices of potentials for which the backflow distribution may exhibit specific features.

\subsection*{Circular polarization}

In the monochromatic case, the transverse component of a real electric field corresponding to circular polarization can locally be written as
\begin{subequations}\label{cipol}
\begin{align}
E_x(\tau, \bm{\xi}) &= A(\bm{\xi})\cos\tau + B(\bm{\xi})\sin\tau, \label{cipolx}\\
E_y(\tau, \bm{\xi}) &= -B(\bm{\xi})\cos\tau + A(\bm{\xi})\sin\tau. \label{cipoly}
\end{align}
\end{subequations}
Here $A(\bm{\xi})$ and $B(\bm{\xi})$ are real amplitude functions that determine the local structure of the transverse field.

Introducing
\begin{equation}
R(\bm{\xi}) = \sqrt{A^2(\bm{\xi}) + B^2(\bm{\xi})},
\end{equation}
the field can equivalently be expressed as
\begin{equation}
\bm{E}_\perp(\tau,\bm{\xi})=R(\bm{\xi})
\begin{pmatrix}
\cos\!\big(\tau+\varphi(\bm{\xi})\big)\\
\sin\!\big(\tau+\varphi(\bm{\xi})\big)
\end{pmatrix},
\end{equation}
where the local phase $\varphi(\bm{\xi})$ is defined (for $R\neq 0$) by
\begin{equation}
\cos\varphi(\bm{\xi})=\frac{A(\bm{\xi})}{R(\bm{\xi})},
\qquad
\sin\varphi(\bm{\xi})=-\frac{B(\bm{\xi})}{R(\bm{\xi})}.
\end{equation}

The squared magnitude of the transverse field,
\begin{equation}\label{polkw}
\bm{E}_\perp^2(\tau,\bm{\xi})=E_x^2(\tau,\bm{\xi}) + E_y^2(\tau,\bm{\xi})=A^2(\bm{\xi}) + B^2(\bm{\xi}),
\end{equation}
is independent of time.

The backflow condition \eqref{lbm1} therefore requires the simultaneous vanishing of both real amplitudes,
\begin{equation}\label{simab}
A(\bm{\xi}_0)=0, \qquad B(\bm{\xi}_0)=0.
\end{equation}
These are two independent real constraints imposed on the transverse plane. In the generic case where $A(\bm{\xi})$ and $B(\bm{\xi})$ are independent functions, their simultaneous zeros form isolated points corresponding to
intersections of the zero sets of $A$ and $B$.

At such points one has $R(\bm{\xi}_0)=0$, so that the transverse electric field vanishes identically and the notion of polarization (specified by the main components of the electric field) is no longer defined there. All statements about circular polarization therefore refer to the local structure of the field in an arbitrarily small neighborhood of these points, where $R(\bm{\xi})\neq 0$.

It is worth emphasizing that the point-like structure of the backflow condition relies on the presence of two independent real amplitude functions. If an additional constraint is imposed, for instance $B(\bm{\xi})\equiv 0$, the local phase becomes fixed (up to a discrete sign of the amplitude), and the backflow condition reduces to a single real equation $A(\bm{\xi}_0)=0$. In such situations, the set of points satisfying the backflow condition in general forms curves, similarly to the case of linear polarization discussed below. This effect is illustrated in Sect. \ref{ce}.

In the generic circularly polarized case, however, the leading-order transverse field vanishes only at isolated points, in the vicinity of which higher-order longitudinal contributions may dominate in finite regions due to field continuity, giving rise to optical backflow.

\subsection*{Linear polarization}

For linear polarization, the transverse components of the real electric field are locally related by a proportionality factor,
\begin{equation}\label{alpha_lin}
E_y(\tau,\bm{\xi}) = \alpha(\bm{\xi}) \, E_x(\tau,\bm{\xi}),
\end{equation}
where the $x$ component can be written as
\begin{equation}\label{linad}
E_x(\tau,\bm{\xi})=A(\bm{\xi}) \cos\!\big(\tau+\varphi(\bm{\xi})\big),
\end{equation}
with $A(\bm{\xi})$ denoting the real amplitude and $\varphi(\bm{\xi})$ the local phase.

Once the local direction of polarization is fixed, i.e.\ the function $\alpha(\bm{\xi})$ is specified (even if it varies from point to point), the two conditions in \eqref{lbm1} become equivalent and reduce to a single real constraint,
\begin{equation}\label{arze}
A(\bm{\xi}_0)=0.
\end{equation}

This single real condition imposed on two transverse coordinates generically defines one-dimensional solution sets, namely curves in the transverse plane. Accordingly, for linear polarization with a well-defined local direction, the leading-order suppression of the longitudinal Poynting vector—and hence the appearance of optical backflow—occurs along lines, in contrast to circular polarization, where it is confined to isolated points.

\subsection*{Radial polarization}

For radial polarization, the transverse electric field oscillates locally along the radial direction in the transverse plane. Introducing polar coordinates $(\xi,\phi)$, with $\hat{\bm{e}}_\xi$ and $\hat{\bm{e}}_\phi$ denoting the radial and
azimuthal unit vectors, respectively, the real transverse field can be written as
\begin{equation}\label{radreal}
\bm{E}_\perp(\tau,\bm{\xi})=E_\xi(\tau,\bm{\xi})\,\hat{\bm{e}}_\xi,
\end{equation}
where
\begin{equation}\label{radamp}
E_\xi(\tau,\bm{\xi})=A(\bm{\xi})\cos\!\big(\tau+\varphi(\bm{\xi})\big).
\end{equation}
The azimuthal component vanishes identically,
\begin{equation}
E_\phi(\tau,\bm{\xi})=0,
\end{equation}
which fixes the local polarization direction at every point in the transverse
plane.

Under these conditions, the backflow requirement \eqref{lbm1} again reduces to the single real constraint
\begin{equation}\label{radzero}
A(\bm{\xi}_0)=0.
\end{equation}
As in the linear case, this condition defines curves in the plane $\xi_z=\mathrm{const}$, so that for radial polarization the leading-order transverse field may vanish along extended lines, enabling higher-order longitudinal contributions to dominate locally and generate optical backflow along one-dimensional sets.

\subsection*{Azimuthal polarization}

For azimuthal polarization, the transverse electric field oscillates tangentially to circles centered on the propagation axis. In polar coordinates, the real transverse field takes the form
\begin{equation}\label{azreal}
\bm{E}_\perp(\tau,\bm{\xi})=E_\phi(\tau,\bm{\xi})\,\hat{\bm{e}}_\phi,
\end{equation}
with
\begin{equation}\label{azamp}
E_\phi(\tau,\bm{\xi})=A(\bm{\xi})\cos\!\big(\tau+\varphi(\bm{\xi})\big).
\end{equation}
The radial component vanishes identically,
\begin{equation}
E_\xi(\tau,\bm{\xi})=0,
\end{equation}
so that the local polarization direction is uniquely defined throughout the transverse plane.

In complete analogy with the radial case, the backflow condition \eqref{lbm1} reduces to the single real requirement
\begin{equation}\label{azzero}
A(\bm{\xi}_0)=0.
\end{equation}
This condition again defines one-dimensional subsets in the transverse plane, implying that optical backflow for azimuthal polarization occurs along curves rather than being localized at isolated points.

In summary, the geometry of optical backflow regions is dictated by the number of independent real constraints imposed by the polarization state: circular polarization leads to isolated backflow points (unless an additional global constraint fixes the polarization phase, in which case extended curves may also occur), whereas linear, radial, and azimuthal polarizations give rise to extended backflow curves in the transverse plane.

It is worth noting that related studies~\cite{staf, bek2} address energy backflow in different formulations of vector fields and asymptotic regimes of the electromagnetic field. In the paraxial framework, radial polarization is defined as a purely transverse, azimuthally symmetric field, and the longitudinal Poynting component does not generally exhibit backflow in this approximation~\cite{bek2}. In the present work, we consider a more complete solution of the near-axis Maxwell equations. In this approach, longitudinal field components and their interference appear at higher order, and the resulting backflow scales as $\varepsilon^4$, which explains why it is not captured in lowest-order paraxial treatments. This mechanism is consistent with localized backflow regions reported in~\cite{staf}, although the ordering of field contributions differs between the approaches.

\section{Concrete examples}\label{ce}

\subsection{Examples with circular polarization}

Let us first consider configurations exhibiting circular polarization, realized by choosing identical scalar potentials,
\begin{equation}
V_+(\bm{\xi}) = V_-(\bm{\xi})= V_R(\bm{\xi}) \equiv V(\bm{\xi}) .
\end{equation}
As a representative paraxial solution, we take
\begin{equation}\label{excircpt}
V(\bm{\xi}) =\frac{\xi_x\xi_y}{(1+2 i\varepsilon^2 \xi_z)^3}\,e^{-\frac{\varepsilon^2(\xi_x^2+\xi_y^2)}
{1+2 i\varepsilon^2 \xi_z}} .
\end{equation}
This potential belongs to the Gaussian--polynomial class and satisfies the paraxial wave equation exactly.

Substituting~(\ref{excircpt}) into the paraxial field expressions~(\ref{parpeeq}) and retaining only the leading $\delta^0$ terms yields the transverse electric field components in the plane $\xi_z=0$,
\begin{align}\label{circptE}
\mathcal{E}_x^{(0)} &= -i \mathcal{E}_y^{(0)} \\
&= \big[-i\xi_x(1-2\varepsilon^2\xi_y^2)+\xi_y(1-2\varepsilon^2\xi_x^2)\big] \,e^{-\varepsilon^2(\xi_x^2+\xi_y^2)} .\nonumber
\end{align}
Away from the zeros of the field, the leading-order transverse electric field thus exhibits a locally circular polarization.

\begin{figure}[h]
\centering
\includegraphics[width=0.45\textwidth]{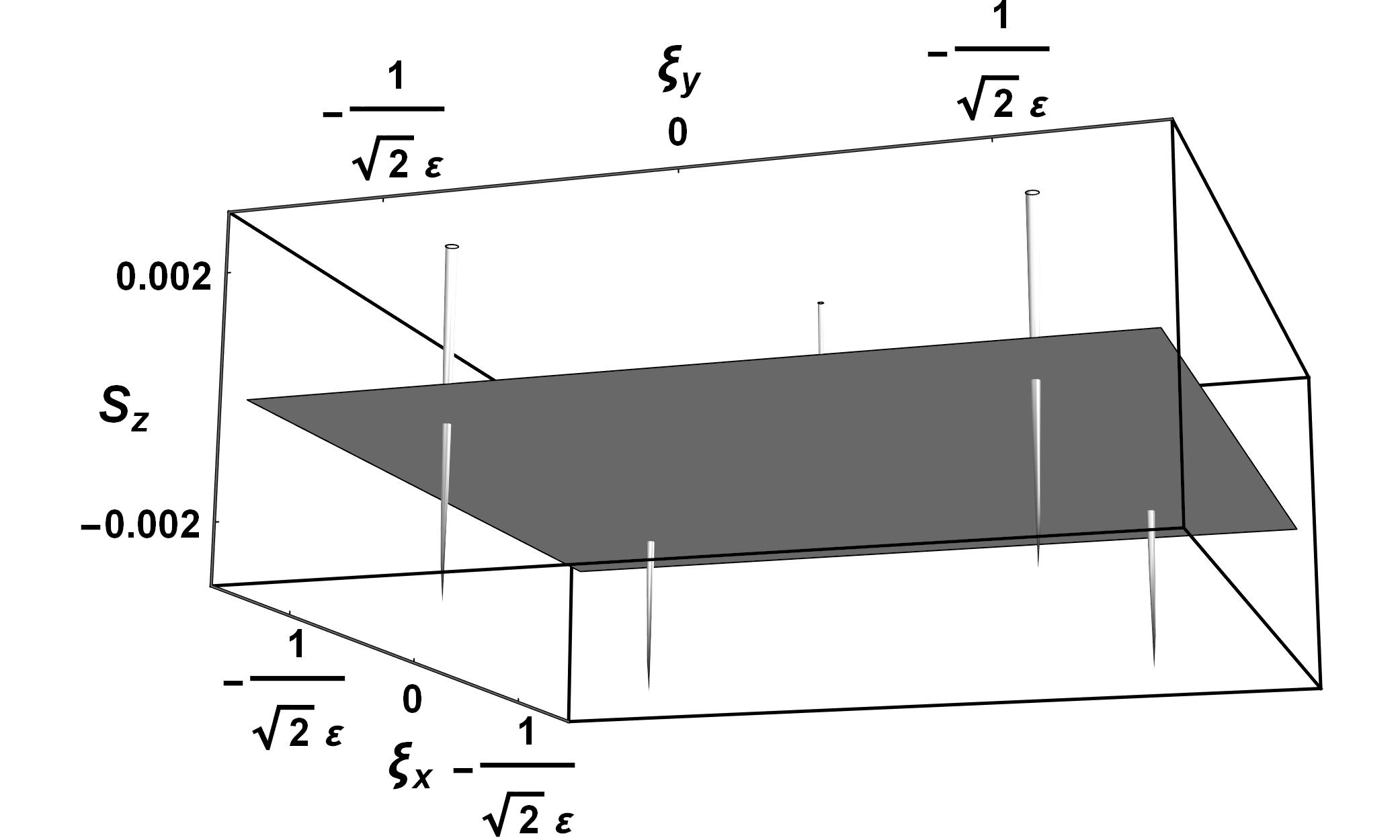}
\caption{Optical backflow for circular polarization, shown for the potentials~(\ref{excircpt}) with $\varepsilon=0.15$ in the plane $\xi_z=0$. The backflow is localized around discrete points in the transverse plane. An additional reference plane at zero is included to indicate explicitly where the longitudinal component of the Poynting vector becomes negative.}
\label{cirback}
\end{figure}

In this case the backflow conditions reduce to
\begin{equation}\label{cocibe}
\xi_x(1-2\varepsilon^2\xi_y^2)=0 \qquad \text{and} \qquad \xi_y(1-2\varepsilon^2\xi_x^2)=0 .
\end{equation}
Their solutions consist of isolated points, namely
\begin{equation}\label{posoli}
\xi_x=\xi_y=0 ,
\end{equation}
together with four additional points,
\begin{equation}\label{posolia}
\xi_x=\pm \frac{1}{\sqrt{2}\varepsilon}, \qquad \xi_y=\pm \frac{1}{\sqrt{2}\varepsilon} .
\end{equation}
An evaluation of the higher-order conditions~(\ref{condpma}) shows that
\begin{equation}
\bm{\nabla}_\perp \partial_{\xi_z} V = 0 \qquad \text{at}\qquad \xi_x=\xi_y=0 ,
\end{equation}
so that optical backflow does not occur at the origin. In contrast, at the points~(\ref{posolia}) the expressions~(\ref{condpma}) are nonvanishing, with moduli equal to $\varepsilon e^{-1}$, leading to genuine point-like optical
backflow, as illustrated in Fig.~\ref{cirback}.

A second example with circular polarization is constructed such that the amplitude $B(\bm{\xi})$ in Eqs.~(\ref{cipol}) vanishes identically. This is achieved by choosing
\begin{equation}\label{excircpta}
V(\bm{\xi}) = V_+(\bm{\xi}) = V_-(\bm{\xi}) = \frac{\xi_x+i\xi_y}{(1+2 i\varepsilon^2 \xi_z)^2}\,e^{-\frac{\varepsilon^2(\xi_x^2+\xi_y^2)}{1+2 i\varepsilon^2 \xi_z}} .
\end{equation}

\begin{figure}[t]
\centering
\includegraphics[width=0.45\textwidth]{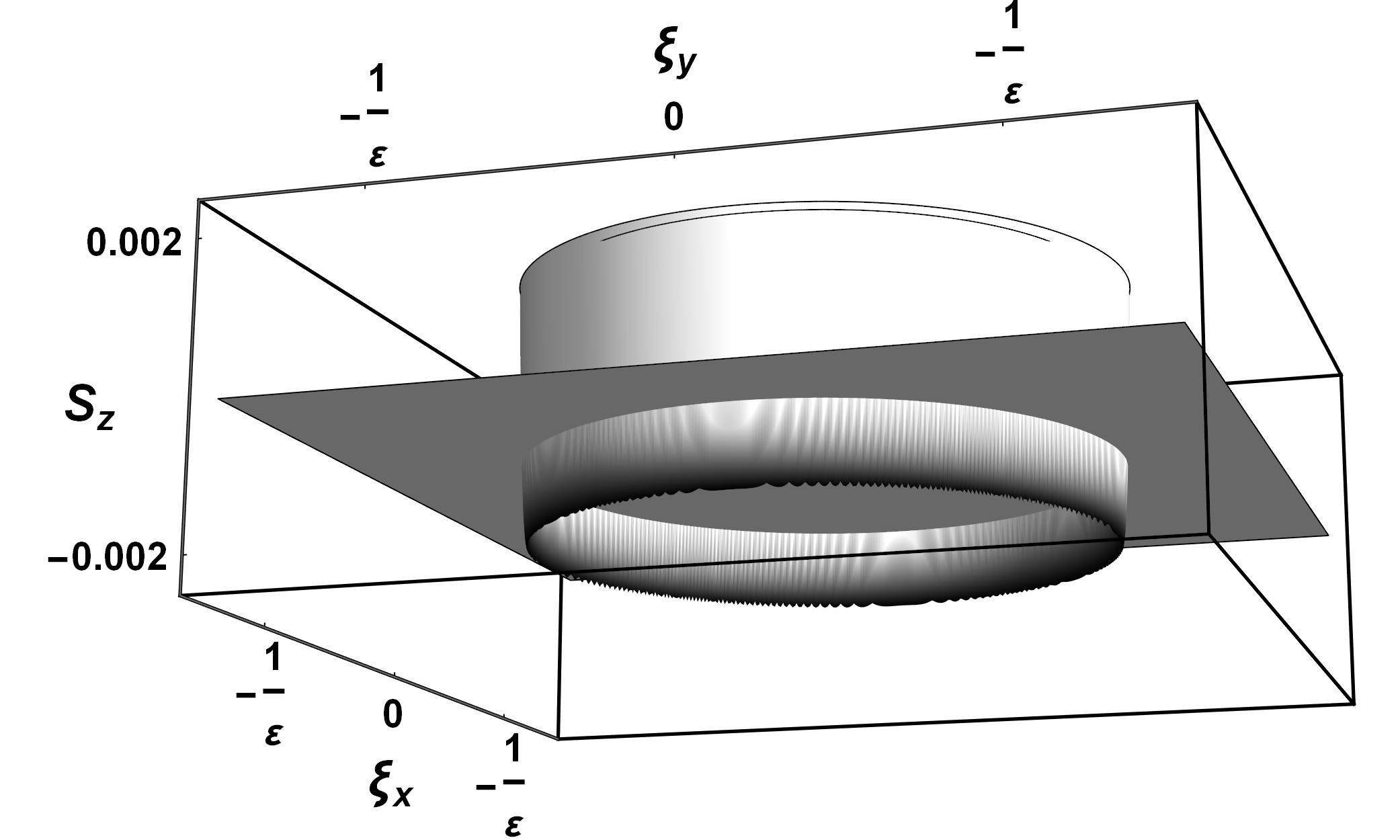}
\caption{Optical backflow for circular polarization, shown for potentials~(\ref{excircpta}) with $\varepsilon=0.15$ in the plane $\xi_z=0$. The backflow is distributed along a circular line in the transverse plane.}
\label{cirback2}
\end{figure}

The corresponding transverse electric field in the plane $\xi_z=0$ setting $\delta=0$ reads
\begin{equation}\label{circptEa}
\mathcal{E}_x^{(0)}=-i \mathcal{E}_y^{(0)}=2\big[1-\varepsilon^2(\xi_x^2+\xi_y^2)\big]
\,e^{-\varepsilon^2(\xi_x^2+\xi_y^2)} .
\end{equation}
The backflow condition now reduces to a single real constraint,
\begin{equation}\label{wgt}
\xi_x^2+\xi_y^2=\frac{1}{\varepsilon^2},
\end{equation}
which defines a circle in the transverse plane, reflecting the axial symmetry of the beam.

The higher-order conditions~(\ref{condpma}) take the form
\begin{subequations}\label{cowali}
\begin{align}
\partial_{\xi_x}\partial_{\xi_z} V_+ + i \, \partial_{\xi_y}\partial_{\xi_z} V_- &=
4i\varepsilon^4 (\xi_x+i\xi_y)^2\big[3 - \varepsilon^2 (\xi_x^2 + \xi_y^2)\big]\nonumber\\
&\times e^{-\varepsilon^2(\xi_x^2+\xi_y^2)},\label{cowalia1}\\
\partial_{\xi_y}\partial_{\xi_z} V_+ - i \, \partial_{\xi_x}\partial_{\xi_z} V_- &=
4\varepsilon^4 (\xi_x+i\xi_y)^2\big[3 - \varepsilon^2 (\xi_x^2 + \xi_y^2)\big]\nonumber\\
&\times e^{-\varepsilon^2(\xi_x^2+\xi_y^2)} .\label{cowali2}
\end{align}
\end{subequations}
The above expressions do not vanish on the circle~(\ref{wgt}), which results in optical backflow distributed along a closed curve, as shown in Fig.~\ref{cirback2}. These two examples demonstrate that, depending on the spatial structure of the paraxial potential, circular polarization may give rise to optical backflow either at isolated transverse points or along extended one-dimensional sets, despite the local rotational character of the transverse electric field away from its degeneracies.

\subsection{Example with linear polarization}

Now we pass to a configuration exhibiting linear polarization, defined by the potentials introduced in~\cite{trvec},
\begin{subequations}\label{exlinx}
\begin{align}
V_+(\bm{\xi}) &=
\frac{i\xi_x}{(1+2 i\varepsilon^2 \xi_z)^2}\,e^{-\frac{\varepsilon^2(\xi_x^2+\xi_y^2)}{1+2 i\varepsilon^2 \xi_z}}, \\
V_-(\bm{\xi}) &=\frac{-\xi_y}{(1+2 i\varepsilon^2 \xi_z)^2}\,e^{-\frac{\varepsilon^2(\xi_x^2+\xi_y^2)}{1+2 i\varepsilon^2 \xi_z}} .
\end{align}
\end{subequations}
Using paraxial field expressions~(\ref{parpeeq}) and retaining only the leading $\delta^0$ contributions, the transverse electric field components in the plane $\xi_z=0$ are obtained as
\begin{subequations}\label{lipolek}
\begin{align}
\mathcal{E}_x^{(0)} &= 2i\big[1 - \varepsilon^2 (\xi_x^2 + \xi_y^2)\big]\,e^{-\varepsilon^2(\xi_x^2+\xi_y^2)}, \\
\mathcal{E}_y^{(0)} &= 0 .
\end{align}
\end{subequations}
This configuration obviously corresponds to linear polarization along the $x$ axis.

The second of the backflow conditions~(\ref{condpm}) is automatically satisfied for the potentials~(\ref{exlinx}). The first condition would in turn require
\begin{equation}\label{ficon}
\xi_x^2 + \xi_y^2 = \frac{1}{\varepsilon^2},
\end{equation}
which coincides with a complete degeneration of the leading-order transverse electric field~(\ref{lipolek}).

The higher-order conditions~(\ref{condpma}) yield
\begin{subequations}\label{cowalib}
\begin{align}
\partial_{\xi_x}\partial_{\xi_z} V_+ + i \, \partial_{\xi_y}\partial_{\xi_z} V_- &=
-4\varepsilon^4 (\xi_x^2-\xi_y^2)\big[3 - \varepsilon^2 (\xi_x^2 + \xi_y^2)\big]\nonumber\\
&\times e^{-\varepsilon^2(\xi_x^2+\xi_y^2)},\\
\partial_{\xi_y}\partial_{\xi_z} V_+ - i \, \partial_{\xi_x}\partial_{\xi_z} V_- &=
-8\varepsilon^4 \xi_x\xi_y \big[3 - \varepsilon^2 (\xi_x^2 + \xi_y^2)\big]\nonumber\\
&\times e^{-\varepsilon^2(\xi_x^2+\xi_y^2)} .
\end{align}
\end{subequations}
These combinations do not vanish simultaneously on the circle~(\ref{ficon}), which ensures the occurrence of optical backflow along this one-dimensional set. Similar behavior was reported for related configurations in~\cite{trback}.

\begin{figure}[h]
\centering
\includegraphics[width=0.45\textwidth]{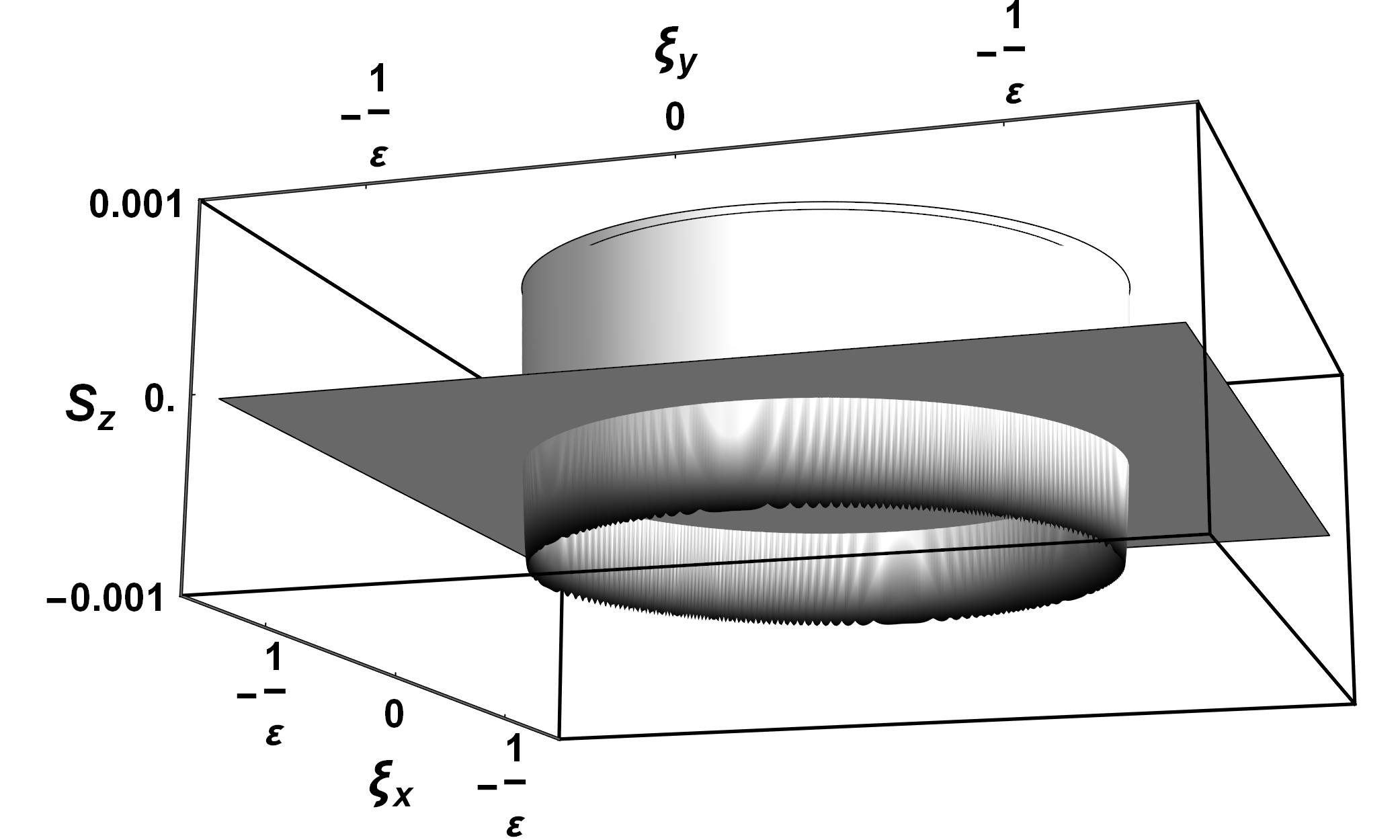}
\caption{Optical backflow for linear polarization, shown for potentials (\ref{exlinx}) with $\varepsilon=0.15$ in the plane $\xi_z=0$. The backflow follows a curve in the transverse plane.}
\label{linback}
\end{figure}

As illustrated in Fig.~\ref{linback}, negative values of the longitudinal component of the Poynting vector form an extended structure following a curve in the transverse plane. This reflects the fact that, for linear polarization, the leading-order transverse electric field vanishes on one-dimensional sets. 

The observed behavior is fully consistent with the local analysis presented above, where the backflow condition reduces to a single real constraint.

\section{Summary and conclusions}\label{concl}

In this work, optical backflow in paraxial Gaussian beams has been analyzed within the exact paraxial Maxwell framework. By decomposing the longitudinal component of the Poynting vector according to its paraxial hierarchy, we have shown that negative energy flux arises exclusively from higher-order longitudinal contributions, whereas the leading-order transverse terms must vanish locally. This provides a clear and physically transparent criterion for the occurrence of optical backflow and clarifies its dependence on the structure of the scalar potentials and their longitudinal derivatives.

A central result of the analysis is that the spatial geometry of backflow regions is governed by polarization through the number of independent local constraints imposed on the transverse field. When the local polarization phase is unconstrained, as in the generic case of circular polarization, the suppression of the leading-order transverse field imposes two independent real conditions, and backflow occurs at isolated points. If an additional global constraint fixes the polarization phase, this reduces effectively to a single real condition, allowing extended one-dimensional backflow curves. In contrast, for linear, radial, or azimuthal polarization—where the phase is locally fixed by symmetry—backflow generically appears along curves.

The dimensionality of the backflow loci follows directly from the number of independent real constraints and is therefore structurally stable under small perturbations of the field. An auxiliary parameter $\delta$ introduced to separate contributions of different paraxial weights serves solely as an ordering device. The fields themselves satisfy the exact paraxial Maxwell equations, and no truncation of the underlying solution is performed. The emergence of negative longitudinal flux thus represents an intrinsic higher-order vectorial effect within the paraxial regime rather than an artifact of asymptotic expansion.

Explicit Gaussian–polynomial examples demonstrate that both point-like and extended backflow structures can be realized entirely within the paraxial domain, without invoking tightly focused or strongly nonparaxial configurations. These results highlight the essential role of vectorial interference beyond scalar theory and provide a systematic framework for understanding and engineering polarization-controlled energy-flow patterns in structured optical fields.

The formalism developed here can be extended to more complex beam families and to regimes in which higher-order or nonparaxial corrections become significant. Such generalizations may elucidate the interplay between field topology, polarization structure, and localized energy transport, and may motivate future experimental investigations of controlled optical backflow.

\end{document}